# Unsupervised seismic facies classification using deep convolutional autoencoder

Vladimir Puzyrev* and Chris Elders

*School of Earth and Planetary Sciences and Curtin University Oil and Gas Innovation Centre, Curtin University, Perth, WA 6102, Australia.*

**Abstract**

With the increased size and complexity of seismic surveys, manual labeling of seismic facies has become a significant challenge. Application of automatic methods for seismic facies interpretation could significantly reduce the manual labor and subjectivity of a particular interpreter present in conventional methods. A recently emerged group of methods is based on deep neural networks. These approaches are data-driven and require large labeled datasets for network training. We apply a deep convolutional autoencoder for unsupervised seismic facies classification, which does not require manually labeled examples. The facies maps are generated by clustering the deep-feature vectors obtained from the input data. Our method yields accurate results on real data and provides them instantaneously. The proposed approach opens up possibilities to analyze geological patterns in real time without human intervention.

*Keywords:* Interpretation, Seismic facies, Deep learning, Convolutional neural network, Autoencoder

## 1. Introduction

Seismic facies analysis plays a key role in the derivation of reservoir properties from seismic attributes (Brown, 2011). Accurate and fast interpretation of facies provides a reference for further analysis of geological conditions. With the widespread use of 3D seismic technology and dramatic growth and complexity of seismic data, manual facies analysis and interpreting geological patterns becomes an extremely time-consuming task. The interpretation results are inevitably affected by the subjectivity of the interpreter. To address these issues, automatic seismic interpretation tools have drawn a lot of attention in recent years with the latest advances in computing and data analysis.

* *Corresponding author, vladimir.puzyrev@gmail.com*

This trend was boosted by the recent meteoric rise of deep learning (DL) models in many fields of science and technology. These methods allow for the detection and exploitation of nonlinear dependencies in the data without specifying a particular model in advance and using hand-engineered features. A particular class of DL models, convolutional neural networks (CNN), have shown tremendous success in image processing, pattern recognition, and object detection systems. In the past few years, deep CNN have been actively applied to various geophysical problems including detection of faults (Araya-Polo et al., 2017; Huang et al., 2017; Wu et al., 2019), seismic facies classification (Dramsch and Lüthje, 2018; Zhao, 2018; Duan et al., 2019), first-break picking (Yuan et al., 2018), seismic horizon picking (Shi et al., 2020) and many others.

Seismic facies classification algorithms can be divided into two major categories, namely, supervised and unsupervised learning techniques. Supervised classification methods (Zhao et al., 2015; Qi et al., 2016; Liu et al., 2018;) involve data labeling, i.e. manual facies interpretation, which is inevitably affected by the differences in the interpreters' knowledge base and experience level. Unsupervised facies classification algorithms are data-driven way to determine the clusters present in the data, unbiased by the interpreter. The most common traditional unsupervised seismic facies classification methods include principal component analysis (PCA) (Dumay and Fournier, 1988), K-means clustering (Coléou et al., 2003; Galvis et al., 2017), and the self-organizing map (SOM) (Matos et al., 2007; Saraswat and Sen, 2012).

Deep neural networks can also be efficiently applied to clustering problems due to their inherent property of highly nonlinear transformation that allows transformation of data with highly complex structure into more clustering-friendly representations (Aljalbout et al., 2018; Min et al., 2018; Chalapathy and Chawla, 2019). In recent years, several unsupervised algorithms based on DL have been developed, in particular, based on the convolutional autoencoder (CAE) neural network architecture that is widely used for image segmentation, denoising purposes, and other problems of image processing. In a geophysical context, CAE has been used for interpolation of missing seismic traces (Wang et al., 2020), attenuation of marine seismic interference noise (Sun et al., 2020), and segmentation of rock images (Karimpouli and Tahmasebi, 2019). In recent years, the same architecture has been marginally used for facies classification applications. Qian et al. (2020) used the deep CAE scheme to extract features from 2D prestack gathers. Somewhat related to it are a group of "semi-supervised" methods that use both labeled and unlabeled data in training; a recent example of such an approach using generative adversarial networks can be found in Liu et al. (2020).

In this study, we explore a deep CAE for unsupervised seismic facies classification. This approach is entirely data-driven, does not require manual data labeling, and provides the result instantaneously. The remainder of the paper is organized as follows. First, we formulate the problem and describe the methodology in Section 2. In Section 3, we describe the architecture of the deep neural networks employed, their training process, as well as the details of the clustering algorithm. The performance of the method is investigated in Section 4 using marine seismic data from the Northern Carnarvon Basin, North West Shelf, Australia. Finally, the last section summarizes the outcomes and points out future research directions.

**2. Methodology**

Seismic data volumes are huge and consist of highly redundant data which motivated previous applications of data reduction algorithms to preserve only the important features of the seismic character (Coléou et al., 2003). This redundancy makes DL methods a natural choice for handling large-scale seismic dataset processing as these methods benefit from massive amounts of data. By exploiting different layers of abstraction, deep neural networks pick up both the low-level and high-level features in data.

Seismic data interpretation can be formulated as an image segmentation problem. Considering a 2D seismic section as a one-channel image, we split it into many "tiles" thus turning the seismic facies recognition problem into an image classification problem. Deep convolutional networks are the state-of-the-art method for such problems. These networks can be viewed as a natural extension of neural networks for processing images and data with a grid-like topology. They achieve high performance in learning filters that represent repeating patterns and extracting the most important features from segments of data regardless of the specific location of these features within the data. Convolutional layers use shared local filters which is beneficial for processing data with a strong local structure (such as seismic data).

A deep CAE is an unsupervised learning method that is able to learn hierarchical feature representations automatically from the input data. The workflow of our CAE-based facies interpretation consists of three stages:

1. CAE training. First, we split the vertical poststack seismic sections chosen as the training data into individual tiles, normalize this dataset and give it as input and output to the autoencoder. During training, the network learns the mapping from the data space to a lower-dimensional latent feature space.

2. Clustering. Once the CAE is trained, we employ the PCA to extract the dominant components of the feature vector which, in turn, are given to the K-means clustering algorithm to generate the facies map on the training data. Another possible way is to apply clustering directly in the latent space, however, this approach may lead to poor performance (Mukherjee et al. 2019).

3. Pattern recognition. Given new seismic data, we generate its deep features using the encoder, extract the dominant components, and find the best matching to generate the facies map. In order to increase the resolution of the resulting facies map, we extract tiles in an overlapping manner along the entire vertical section. Finally, we postprocess the obtained result to suppress the noise and remove spatially tiny structures.

The method delivers the results in real time (the third stage takes several seconds using a single GPU). Details of the implementation using open-source libraries TensorFlow (Abadi et al. 2016) and scikit-learn (Pedregosa et al., 2011) are described next.

## 3. Implementation details

*3.1 Convolutional autoencoder*

An autoencoder is an unsupervised learning method, which is based on training the neural network to approximate the data by itself via a bottleneck structure (Masci et al., 2011). Unlike traditional supervised DL methods, it does not require large amounts of labeled training examples and can automatically learn discriminative features in data. Various autoencoders have been applied for clustering and anomaly detection tasks in recent years (Guo et al., 2017; Ghasedi et al., 2017; Min et al., 2018; Chalapathy and Chawla, 2019). Convolutional autoencoders (CAE) are a special type of autoencoder that uses convolutional layers to extract high-level features from data while preserving local relations using the convolution kernel in each layer. This makes CAE a natural choice of autoencoder for processing of images and data with local spatial connectivity such as seismic data.

An autoencoder consists of two major parts, the encoder $\mathbf{E}_\phi$ that compresses the input $\mathbf{x}$ to lower-dimensional features $\mathbf{h}$ and the decoder $\mathbf{D}_\theta$ that takes the latent features as input and reconstructs the original data as closely as possible:

$$\mathbf{h} = \mathbf{E}_\phi(\mathbf{x}), \tag{1}$$

$$\tilde{\mathbf{x}} = \mathbf{D}_\theta(\mathbf{h}). \tag{2}$$

Choosing the architecture of both the encoder and decoder as a deep CNN allows it to learn hierarchical feature representations by exploiting deep features in the input data. This process of encoding can be viewed as a projection of the higher-dimensional input data onto a lower-dimensional space. The encoder, in other words, is a nonlinear function that allows automatic extraction of feature representations. The number of hidden features should be sufficient to describe enough variability of the data and thus be able to reveal details of the underlying geologic features. In the decoder, we use upsampling layers in a symmetric manner to restore the original data. The bottleneck structure of the CAE allows it to capture the most important features associated with the input seismic images in the hidden feature layer $\mathbf{h}$. This layer has much smaller dimensionality than the original data, which leads to the creation of a compressed set of information in **h** from which the original data **x** is restored through linear and non-linear relationships. In other words, training of the CAE creates in **h** a more cost-effective representation of **x**.

Figure 1 shows the architecture of the CAE used in the following examples. It is composed of 44 layers: 22 convolutional, batch normalization, and pooling layers in the encoder to build a deep representation of patterns in data, a feature layer, and 21 convolutional, batch normalization, and upscaling layers in the decoder. Each of the layers in the encoder and decoder networks contains between 32 and 256 filters for detecting various features in the input data. The number of filters used in convolutions increases by a factor of 2 after each pooling layer to enable the network to better learn features at higher abstractions, make the representation approximately invariant to small translations of the input, and also reduce the overall number of free parameters. Due to this hierarchical structure when the features extracted at the previous level become the input at the next level, the network is able to analyze the data at different scales. Batch normalization (Ioffe and Szegedy, 2015) is used after the convolutional layers for improving the training performance and regularization purposes. Rectified linear units (ReLU) are chosen as activation functions. The main advantages of these activation functions are their nonsaturating nonlinearity that often works well in training and low computational complexity. Based on a series of experiments (Puzyrev, 2019), we choose leaky rectified linear units (ReLU) as activation functions of the CAE. Two different sizes of the convolution kernel, namely 3 × 3 and 5 × 5, were tested in the model as described below.

Like any other neural network, CAE needs to be trained. The training, however, does not require labeled examples and is performed using the input data as the output data, namely

$$\tilde{\mathbf{x}} = \mathbf{D}_\theta \left( \mathbf{E}_\phi \left( \mathbf{x} \right) \right) = \mathbf{x}. \tag{3}$$

This allows the expensive procedures of manual data preparation to be avoided. The weights and biases of the encoder and decoder networks, $\phi$ and $\theta$, respectively, are iteratively updated during training by minimizing the reconstruction error. While the training process is time-consuming for large datasets and deep networks, once it is finished, the network can be used at a very low computational cost.

*3.1.1 Loss functions*

The role of the loss function is to ensure that the learned representation preserves important features of the initial data. A rather common choice for regression problems is the mean squared error (MSE). This loss is also commonly used in denoising autoencoders as the distance measure between the original and decoded data. However, a custom loss function instead of the standard MSE can improve the performance of an autoencoder (Creswell et al. 2017). Cross-entropy losses are preferred over MSE for classification problems. Here, we use the following loss function which is a weighted sum of the MSE over the normalized training dataset and the binary cross-entropy of the binarized data:

$$L(x,\tilde{x}) = L_{MSE}(x,\tilde{x}) + \alpha L_{BCE}(x_b,\tilde{x}_b). \tag{4}$$

The MSE and BCE losses are defined as follows

$$L_{MSE}(x,\tilde{x}) = \frac{1}{N}\sum_{i=1}^{N}(x_i - \tilde{x}_i)^2, \tag{5}$$

$$L_{BCE}(x,\tilde{x}) = -\frac{1}{N}\sum_{i=1}^{N}\left(x_i \cdot \log \tilde{x}_i + (1-x_i) \cdot \log(1-\tilde{x}_i)\right); \tag{6}$$

and the subscript $b$ denotes the binarized data. The role of the BCE loss is to penalize the structural difference between the original and reconstructed data. Nesterov-accelerated adaptive moment estimation (Nadam) algorithm (Dozat, 2016) is employed. Table 1 reports the parameters of the networks and the minimum error (4) achieved on the training dataset. Both networks achieve similar accuracy on the training data, although the second one takes significantly longer to train due to the higher number of training parameters and more epochs required to reach the error plateau. At the same time, the facies maps generated with Model 2 are slightly smoother and contain less "noise" and thus we employ this network in the following numerical examples. Once the training is complete, the input data are split into several groups (clusters) as described in Section 3.3.

*3.2 Data*

We train the network on a high quality public domain 3D seismic data acquired in the Northern Carnarvon Basin, Australia's premier hydrocarbon province. This area has been extensively studied, not only in terms of hydrocarbon prospectivity, but also in terms of the structural, stratigraphic, sedimentological and geodynamic evolution of the continental margin. The Northern Carnarvon Basin developed during several successive phases of extension from the Late Carboniferous until the Early Cretaceous, followed by passive margin thermal subsidence to the present day. This has resulted in the accumulation of sedimentary sequences up to 15 km in thickness which were deposited in a variety of different sedimentary environments and hence are represented by a number of different sedimentary and seismic facies (I'Anson et al., 2019; McHarg et al., 2019).

The hc2000a 3D survey was acquired in 2000 and consists of 2324 in-lines and 4470 cross-lines with a 25 m inline spacing and a 12.5 m cross-line spacing covering the northern part of the Exmouth sub-basin and the adjacent Exmouth Plateau. The record length is 9000 ms with a 4ms sample rate. The data are time migrated. We extract 60 2D vertical sections from the available 3D volume, which contain a range of seismic facies with different degrees of reflector continuity, frequency and amplitude response. Data preprocessing is a crucial step in DL applications which may seriously affect the final result. In this particular case, we normalize the data over the entire training dataset to avoid numerical problems in network training since the range of amplitudes is sufficiently large and such variations in different samples may cause errors in gradient updates.

Each vertical seismic section is split into non-overlapping tiles of 96 traces and 48 samples. This size was determined by a series of experiments. Smaller sizes make classification of individual tiles more challenging while larger tiles increase the chances of having several facies in one tile. 54 regularly spaced in-lines and cross-lines are used as the training data, which constitutes a sufficiently large and representative set of training data for a deep CAE to allow it to achieve high accuracy and generalization, i.e. perform meaningful classification of facies both for the data used in training (training set) as well as for new, previously unseen data (test set). Figure 2 shows the level of CAE reconstruction for ten sample tiles representing various geological facies. This indicates that the main reflector geometries are reconstructed with a high degree of accuracy, while the small amount of noise present in the original data is absent in the CAE output.

*3.3 Clustering*

Once the network is trained, we need to cluster the deep-feature vectors obtained from the input images for a further construction of a facies map. In order to reduce the dimensionality of the feature vector **h**, we employ the PCA for extraction of the dominant components. In Figure 3, we illustrate the contribution of each PCA component into the total explained variance for Models 1 and 2. The first 20 components in these two cases explained, respectively, more than 49% and 47% of the total variance.

The obtained PCA components are used for clustering of the training samples. Clustering can be done by various unsupervised methods including K-means clustering, fuzzy c-means, or self-organizing maps. Without losing generality, to generate the facies maps in this study, we apply a K-means clustering algorithm (Hartigan and Wong, 1979) based on the distance between the principal components of the latent features extracted from the training data. Each of the $96 \times 48$ tiles is classified into one of several classes based on the similarity of their principal components. In order to increase the resolution of the resulting facies classification, we employ a sliding window over the entire vertical seismic section in an overlapping manner. Figure 4 illustrates this concept for the case when the fine grid is four time finer than the original grid in each dimension. For each sliding window position, we run the algorithm to determine its facies class and add this class to the 16 fine cells contained inside the window using the weight matrix shown in Figure 4b. The facies class for the fine cell is determined as the one having the highest weight.

**4. Numerical results**

In this section, we show the facies classifications generated for the considered Northern Carnarvon Basin seismic data. Figure 5a shows the CAE interpretation of an inline section from the training dataset. Seismic facies appearing in this dataset are automatically divided into 5 classes. Tiles with no reflections (water) are determined precisely. The relatively high amplitude and laterally continuous passive margin sequences (Muderong Shale and younger; red color) are distinguished from the continuous but lower amplitude reflectors of the Barrow Group (yellow color). Incisions within the passive margin sequence are also recognized. The Barrow Group sediments are in turn distinguished from diffuse reflections in the underlying Upper Jurassic sequence (green color). However, the algorithm has failed to distinguish between the faulted and rotated Triassic sequences forming the basin flank, and the more seismically transparent facies forming the basin fill. Some high amplitude events, which may represent igneous intrusions, are also picked out. In order to obtain a cleaner result, we also postprocess the algorithm output using a combination of morphological closing and median filtering techniques, which allows to get rid of spatially tiny structures (Figure 5b). For comparison, we also show the results of manual interpretation of the same inline section in Figure 5c.

The distinction between the different formations is also clearly apparent in the cross lines, as shown in Figure 6 for a cross-line section from the test dataset. The angular unconformity between the Barrow Group sediments (yellow) and the overlying passive margin sequence (red) is clear, although there is some "leakage" of the facies below the unconformity. Once again, there is little discrimination between faulted units and the rift fill, but high amplitude igneous intrusions are clearly recognizable (purple).

## 5. Discussion

Deep learning methods have attracted considerable interest from the geophysical community and made artificial intelligence one of the main focuses of attention from both academia and industry. Automatic methods for seismic data analysis are commonly applied nowadays for recognizing geologically meaningful patterns and identifying various features such as faults, sequence boundaries, and unconformities. Application of deep learning methods in seismic facies interpretation could significantly reduce the manual labor and subjectivity of a particular interpreter present in conventional methods. Unsupervised learning methods are of particular interest nowadays and their role in seismic interpretation is expected to grow rapidly in the near future.

By employing the proposed CAE architecture, deep features are learned from poststack seismic images in an unsupervised way thus eliminating the need of labeling training data and avoiding interpreter's subjectivity in definition and delineation of seismic facies. While the lack of need for labeled training data is a benefit of unsupervised methods, this also limits our possibilities to control the final result. Although the results compare favorably with the manual interpretation, there are still some obvious differences. The tiles are considered individually, while manual facies interpretation would heavily rely on their positions relative to each other. Future research may explore adding additional attributes such as frequency variation with depth to the training data. Another direction for future research is the extension of the method to 3D.

## 6. Conclusions

We applied a deep convolutional autoencoder to automatically extract the dominant features in seismic data and classify this data into various seismic facies. The method successfully extracted the dominant features of the seismic data in the training and test sets and classified them into five seismic facies. The workflow consists of three stages: training of the CAE, clustering of the training data, and recognition of seismic patterns in new data. Facies maps are generated by applying the K-means clustering algorithm on the principal components of the compressed features. Normalization of input data and fine-tuning the network architecture has an important effect on the clustering results. The first stage of the workflow is computationally expensive, since the network is trained on large datasets in order to learn the underlying dependencies in data and be able to generalize to new inputs. However, the first and second stages are only performed once. Processing of new data is performed instantaneously and, as demonstrated by the examples in this study, allows for sufficiently accurate estimation of the basic seismic facies. We also note that the trained network can be adapted to other datasets by employing the concept of transfer learning and fine tuning the network on the new data to improve training efficiency.

**Acknowledgements**

The authors acknowledge support from the Curtin University Oil and Gas Innovation Centre (CUOGIC) and the Institute for Geoscience Research (TIGeR). This work was supported by resources provided by the Pawsey Supercomputing Centre with funding from the Australian Government and the Government of Western Australia.

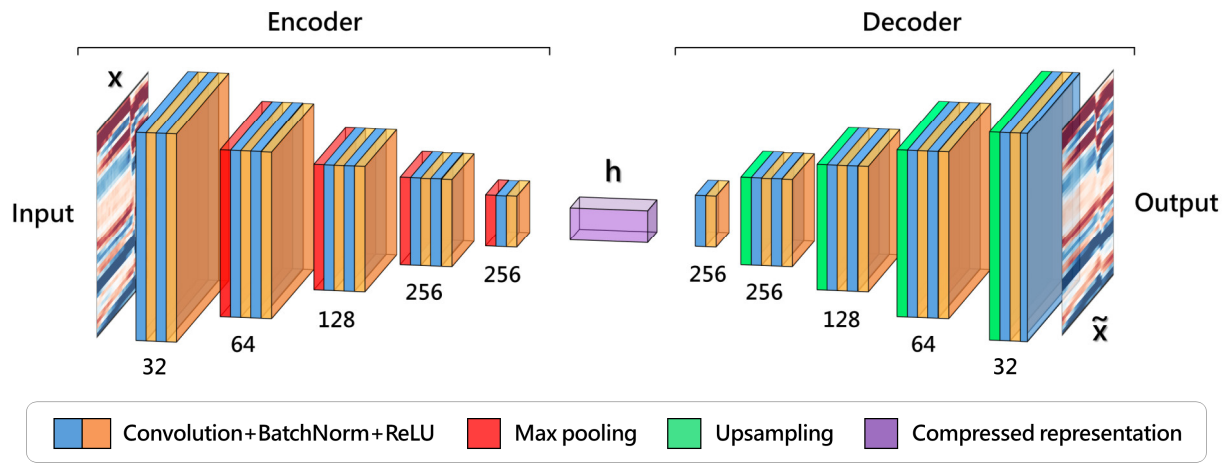

**Figure 1.** Architecture of the CAE. Color rectangles denote multichannel feature maps. The number of channels is shown at the bottom.

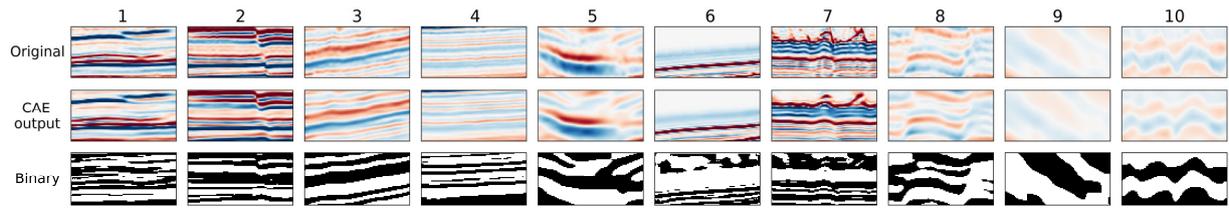

**Figure 2.** Examples of 96 × 48 tiles: original input data $\mathbf{x}$ (top row), CAE-reconstructed data $\tilde{\mathbf{x}}$ (middle row), and binary maps used in the loss function (4) (bottom row).

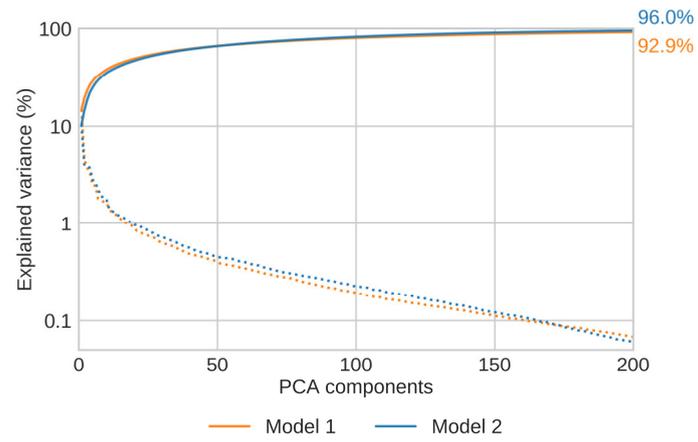

**Figure 3.** Sum of the explained variance (solid lines) and its individual contributions (dotted lines) for varying number of PCA components.

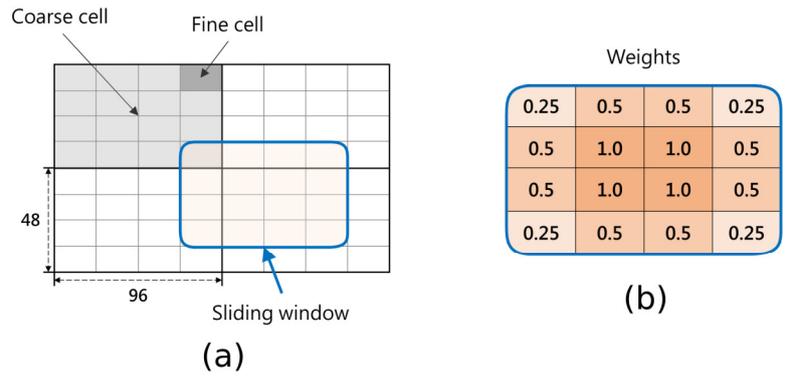

**Figure 4.** 4x coarse-to-fine grid interpolation scheme. Left: sliding window example. Right: weight matrix.

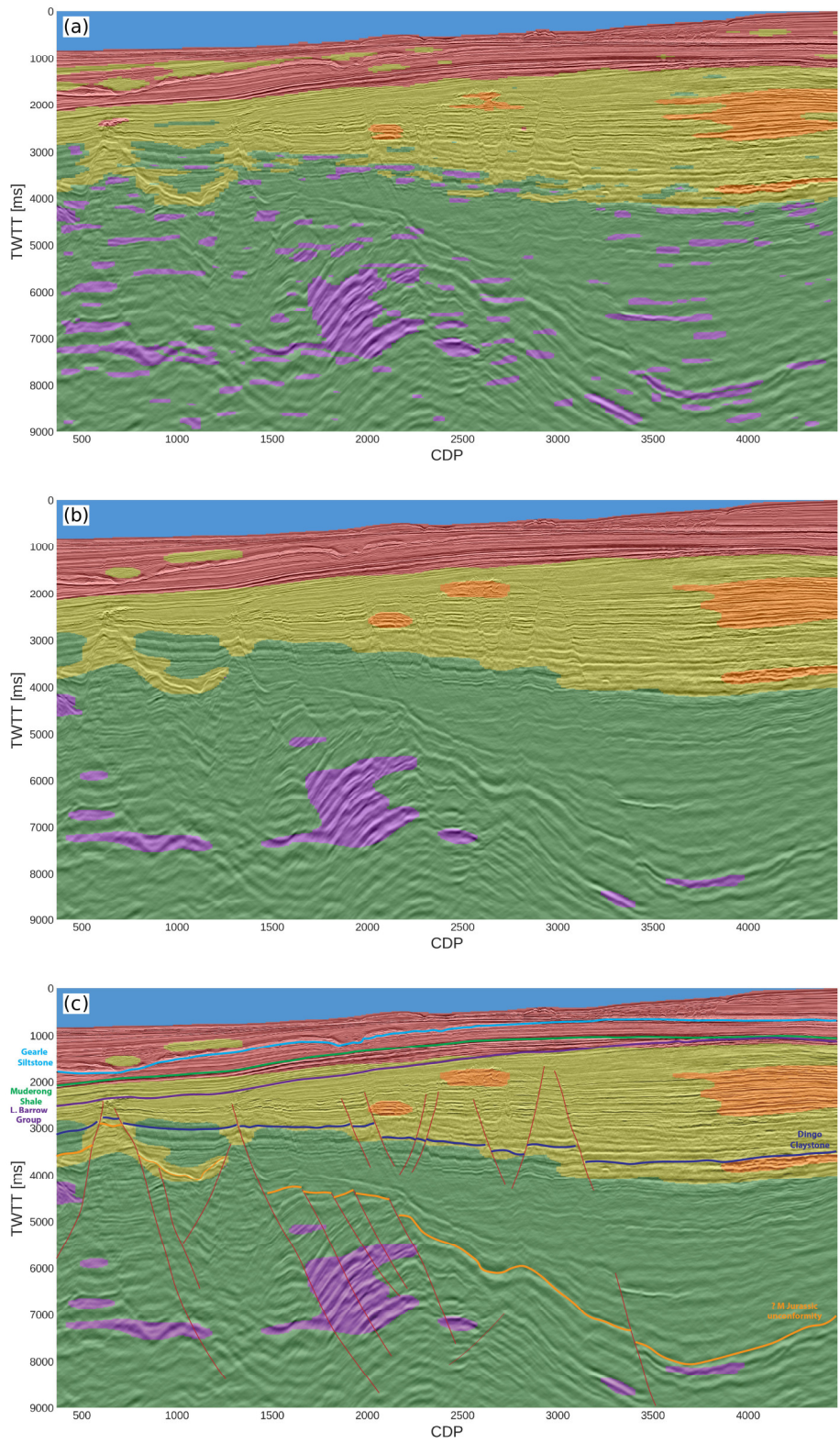

**Figure 5.** Results of the CAE facies interpretation of a vertical seismic section from the training dataset. Top: original network output (a). Middle: automatically filtered image (b). Bottom: manual interpretation (c).

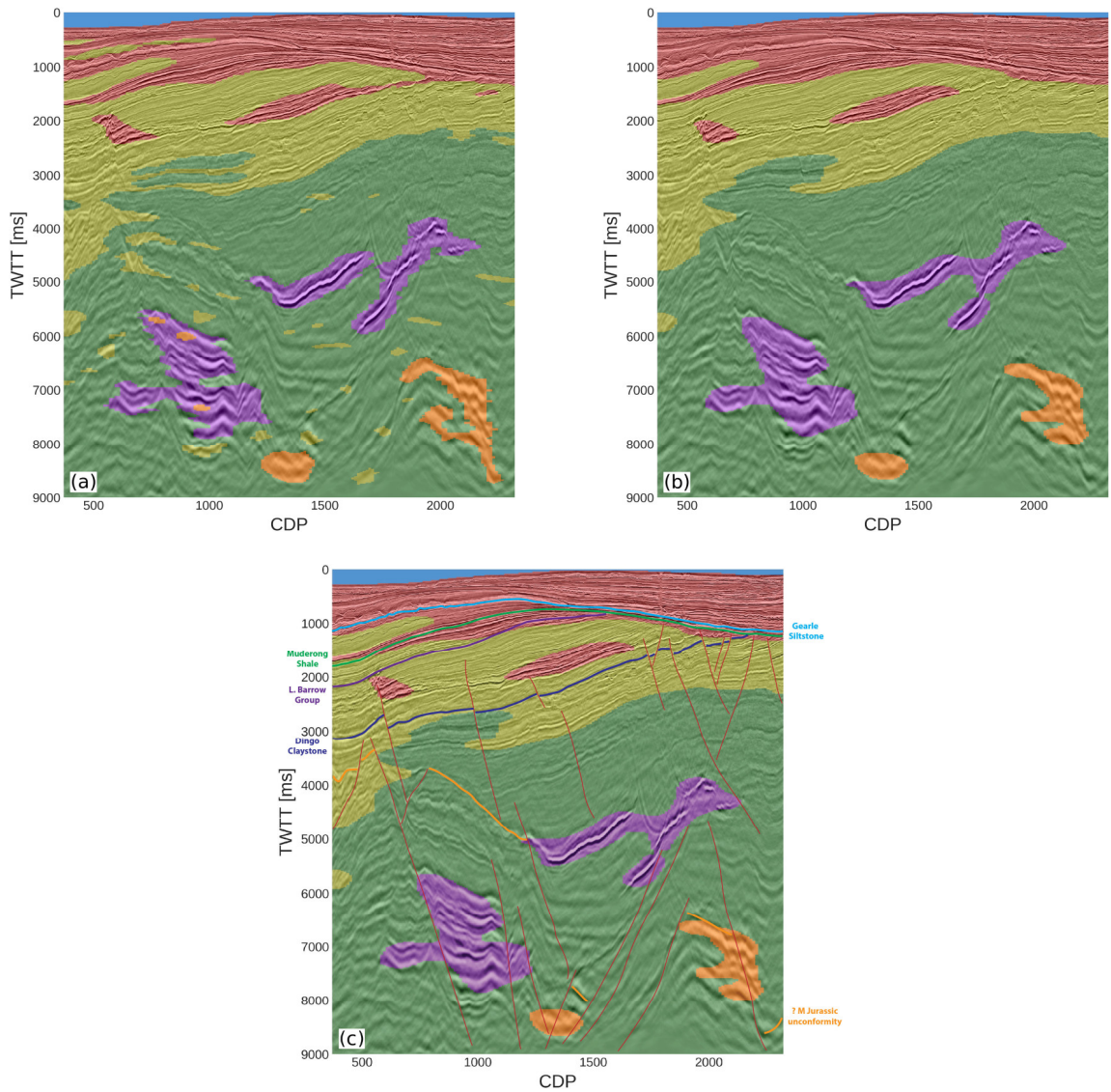

**Figure 6.** Results of the CAE facies interpretation of a vertical seismic section from the test dataset. Top left: original network output (a). Top right: automatically filtered image (b). Bottom: manual interpretation (c).

| Model | Kernel size | Trainable parameters | Minimum training error |
|---|---|---|---|
| **1** | 3 × 3 | 4.7 M | $9.6 \times 10^{-3}$ |
| **2** | 5 × 5 | 13.1 M | $9.5 \times 10^{-3}$ |

**Table 1.** Parameters of the networks and the lowest training set errors using loss (4) with $\alpha = 0.005$.